# Offering A Product Recommendation System in E-commerce

**Ruma Dutta[1,3], Debajyoti Mukhopadhyay[2,3]**

[1] Netaji Subhash Engineering College, Garia, Kolkata 700152, India
rumadutta2006@gmail.com
[2] Maharashtra Institute of Technology, Pune 411038, India
debajyoti.mukhopadhyay@gmail.com
[3] WIDiCoReL, Green Tower, Block C, Flat 9/1, Golf Green, Kolkata 700095, India

**Abstract.** This paper proposes a number of explicit and implicit ratings in product recommendation system for Business-to-customer e-commerce purposes. The system recommends the products to a new user. It depends on the purchase pattern of previous users whose purchase pattern is close to that of a user who asks for a recommendation. The system is based on weighted cosine similarity measure to find out the closest user profile among the profiles of all users in database. It also implements Association rule mining rule in recommending the products. Also, this product recommendation system takes into consideration the time of transaction of purchasing the items, thus eliminating sequence recognition problem. Experimental result shows for implicit rating, the proposed method gives acceptable performance in recommending the products. It also shows introduction of association rule improves the performance measure of recommendation system.

**Keywords:** data mining, information retrieval.

## 1 Introduction

World Wide Web makes the life of customers simpler by introduction of e-commerce where commercial activities can be done from own location. Most customers like to have a recommendation system by which customers can see the feedback from other users who already purchased the products. This need gives rise to the demand of a product recommendation system. E-commerce makes use of recommendation systems in one step ahead which not only shows the feedback from other users but also suggests interesting and useful products to customers. They provide consumers with information that is intended to support their recommendation activities. Recommendation systems research is mainly motivated by the need to cope with information overload, lack of user knowledge in a particular domain. There are two types of widely used product recommendation techniques. They are Automated Collaborative Filtering (ACF) and Case-Based Reasoning [2].

The first one, ACF [1] is widely used as the technique for product recommendation in an online store. This approach is based on the feedback given by the previous customers and their feedback is used to recommend the product to a new customer. For example, let us suppose there are three customers 1, 2 and 3. They all have purchased A, B, C products. Additionally, customer 1 and 2 purchased the products D and E. As customer 3 has common interest with customers 1 & 2, the products D and E are recommended for customer 3. Some ACF system can provide the reason and data behind recommendation. Most ACF recommendation system use the formula given in equation (i) which is also known as **mean squared difference formula.**

$$\delta_{UJ} = \frac{1}{\#n} \sum (U_f - J_f)^2 \quad \text{(i)}$$

This formula is used to calculate the similarity between two persons U and J, in terms of their interests on a product. $U_f$ and $J_f$ are the ratings of U and J on the feature f of the product. This paper proposes a different measure which combines the user ratings as well as importance of the items in the particular e-commerce site to identify the closest user whose purchase pattern is closest and significant for the user for whom item is to be recommended.

ACF approaches can be classified as non-invasive and invasive approaches, based on how the user's preferences are recorded in an ACF system [1][3]. In invasive approach, user's ratings are floating numbers between 0 and 1. Non-invasive approach, the preferences are Boolean values i.e. 0 & 1. For an example, there are four products P1, P2, P3 and P4. User 1 has used the products P1, P2, P4. In non-invasive approach, the ratings would be 1, 1, 0, 1. In invasive approach it may be 0.2, 0.5, 0 and 0.6. The values 0 indicate that the User 1 has not rated the product P3. Obvious problem with non-invasive approach is, the product P1 is rated low by User1 but user's rating will be 1 for P1. For this reason, the non-invasive approaches require feedback from more users to recommend any product.

ACF system can be classified as explicit and implicit types. The aforesaid method is explicit ACF where users rate the items. Sometimes rating of the products can be implicit where rating can be derived from the system itself like from user profile recorded during registration.

This paper proposes a hybrid recommender system based on ACF and Association Rule mining, taking into consideration the sequence of purchase. Using association rule derived from transactions, this system recommends more than one product. This recommendation can result into some products which are not available in the recommendation system database. The proposed system also eliminates the problem of sequence recognition. This paper also proposes an implicit rating system which is used for new user who did not use the e-commerce site before.

## 2  Our Approach

Our recommendation system emphasizes four key areas. These are

- Introducing Cosine Similarity measure where rating of the products by users as well as their frequency in transaction database being considered in ACF which is dealt in section 2.1
- An implicit rating system from the transaction database has been introduced in section 2.2
- Section 2.3 discusses about Eliminating sequence recognition problem
- Recommending more than one product using Association rule, dealt in section 2.4

### 2.1  Weighted Cosine similarity measure in ACF

Cosine similarity measure is widely used in information retrieval [5]. Every document is looked upon as vector. The frequencies of terms are the components of vectors. Then Cosine Similarity (equation (ii)) is the process by which the similarity between two document vectors is measured [6]. The Cosine

$$\mathrm{Cos}(d1,d2) = \frac{dot(d1,d2)}{\|d1\| \|d2\|} \qquad \text{(ii)}$$

where,

$dot(d1,d2) = d1[0]*d2[0]+d1[1]*d2[1]+\ldots\ldots$

$\|d1\| = \mathrm{sqrt}\ (d1[0]\wedge 2+d2[0]\wedge 2\ldots\ldots\ldots\ldots..)$

Same Cosine Similarity can be used in ACF. The Product rating of each user can be viewed as a vector and for a user the rating of each product is the component of the vector. Let us suppose, there are two users U1 and U2. There are five products P1, P2, P3, P4, P5. The rating of User1 is 0.5, 0.6, 0, 0.7, 0.8. The vector of User1 is represented as (5, 6, 0, 7, 8). Similarly for User2 whose rating is 0.5, 0.6, 0.6, 0.2, 0.9 will be represented by vector (5, 6, 6, 2, 9). The User3 had purchased first three products P1, P2, P3 and ratings for those products are .4,.5,.6. The User3 vector will be represented as (4,5,6). Now, product has to be recommended for User3. As User3 purchased only P1, P2, P3, the vector considered for User1 will be (5, 6, 0) and User2 will be (5, 6, 6). Now the Cosine Similarity value for User 1 and User 3 is .73 and for User2 and User3 is .99. So User2 is more similar to User3 than User1 and the product P5 is recommended. As User 2 has rated low for product P4, this product is not recommended for User3.

This method can be further modified by introducing the frequency of the items purchased by the user with the rating given by the user for a particular item. The rational behind introducing frequency of items is that, more satisfied the user is with the system, more purchase will be done by the user. This rating is done implicitly, so we call the rating done by purchase pattern of the users as implicit rating.

Let us consider item frequency of the user u for the product i as IF(u,i)

where, IF is Item Frequency ,

u is the user,

i is the item.

IF(u,i) can be defined by two methods

**Method 1**

$$IF(u,i) = n(u,i) / \sum_{I} n(u,I)$$

Where, n(u,i) is the frequency of item i by user u.

Now, if the rating of item P1 by User1 is 0.5 and frequency of purchase of P1 by User1 is 5 and purchase of all products <P1….P5> is 10, then IF(User1,P1) is .5*5/10=.25, similarly for other products. These can be represented as vectors as before.

**Method 2**

$$IF(u,i) = n(u,i) / \max(u,I)$$

Where, n(u,i) is the frequency of item i by user u.

These two methods have been compared in the experimental section of this paper.

### 2.2 Vector Space model for implicit user rating

To recommend the product for users, implicit user rating procedure can be used. Vector space model, a widely used information retrieval model is used for this purpose. Here, two terms item frequency (IF) and Inverse item frequency (IIF) are used. These two terms are defined as follow:

$$IF(u,i) = n(u,i)$$

Where, n(u,i) is the frequency of the item i purchased by the user u

$$\text{IIF(i)}=\log\frac{1+U}{U_i}$$

Where, $U_i$ is the no. of users who purchased the item i

U is total no. of users.

Then the coordinate of user u in axis i can be defined as

$$u_i=\text{IF}(u,i)\text{IIF}(i)$$

All the users can be defined in this way. Then the cosine similarity measure as explained in section 2.1 is used to find out the closest user of the target one.

For new user, Item frequency of item using the formula $\text{IF(i)}=\log\dfrac{U_i}{1+U}$ has been used where U and $U_i$ has its usual meaning as in IIF in vector space model.

## 2.3 Sequence Recognition Problem

Let us consider the following situation. User 1 bought the books Physics-1, Physics-2, Physics-3, Physics -4 from an ACF system. The user bought the books on above order and gave a higher rating on these books. Another user User 2 bought the product Physics 3 and Physics 4. From the system described in 2.1 the User 2 whose rating is similar to User1, will recommend Physics 1 and Physics 2. But this recommendation is of no use as practically, the probability of buying Physics-1 and Physics-2 after buying Physics-3 is very less and also Physics-3 is not purchased after Physics-1 and Physics-2 in the past. This type of problem is very common in ACF and is known as sequence recognition problem.

To eliminate this problem the sequence of purchase has to be recorded. This paper solves this problem by finite state machine. From Section 2.1, User3 has purchased P1,P2,P3 and recommended product is P5. If P5 is not purchased after purchasing P1, P2, P3 then that product is not recommended. This can be represented by the following Finite State Machine.

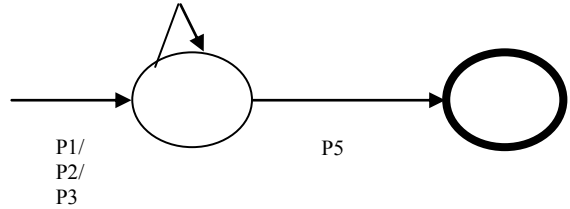

**Fig. 1.** Finite State Machine

## 2.4 Association rule and its use in Product Recommendation

### 2.4.1 Definitions

**Definition 1 – Support:** A transaction t is said to support an item $l_i$, if $l_i$ is present in t. t is said to support a subset of items $X \subseteq A$, if t supports each item l in X. An item set $X \subseteq A$ has a support s in T, denoted by $s(X)_T$, if s% of transactions in T support X.

**Definition 2 – Association Rule:** For a given transaction database T, an association rule is an expression of the form X=> Y [4], where X and Y are subsets of A and X=> Y hols with confidence τ if τ% of transactions in D that support X also support Y. The rule X=> Y has support σ in the transaction set T if σ% of transactions in T support $X \cup Y$.

For recommending more than one product, we can make use of association rule. From table 1 it could be concluded that if P2 is recommended, then P4 and P1 can be recommended also as P2 is bought together with P1 and P4 in most of the cases.

**Table 1.** Transaction IDs vs. Items

| TID(Transaction ID) | Items |
|---|---|
| 100 | P1, P2 |
| 200 | P1,P2,P4 |
| 300 | P1, P4 |
| 400 | P5,P4 |
| 500 | P1,P5 |

Algorithm 1 contains the steps necessary to recommend the products

**Algorithm 1: Recommendation of product**

  Input: D= {Pl1, Pl2,….Plk}  // Database of products
    Plr            // The rating on the products of the User for whom the recommendation is being carried out
    RatD        // Rating of all users on the products of D
    s             // Support
  Output : Recommended products

Step 1: Use Vector Space model on D, Plr and RatD to find the users of the same clusters of the user for whom the recommendation is to be made.
Step 2: Find out the ratings of the products rated best for each of the 5 users for the products which are not present in Plr and store these into plr1 array of size 5..
Step 3: for i=1 to 5
     Loop
Step 4: Take temp_reco= i.
Step 5: Find out whether the product has been bought after any of the products in plr
Step 6: then recommend the product and insert into reco[i]
Step 7: else go to Step 3 and execute Step 4 thru Step 7.
     End Loop
  Step 8: for i= 1 to 5
     Loop
  Step 9: Sort D on the frequency of the products in transactions
Step 10: Find all Association rule using FP-growth algorithm for the product in plr[i] and insert into Reclist1.
  Step 11:  Find out whether the products in Reclist1 bought after any of the products in plr.
  Step 12:  If Step 11 is satisfied, then include that product in plr1[i].
  Step 13: Else go to Step 8.
     End Loop
  Step 14: End.

## 3  Experimental Result

Experiments have been carried out using the dataset, generated by a survey. The products are in book domain.  Users of 4 classes give their rating on 60 items. 80% users are in training dataset and 20% users are in test dataset.  Transactions of these items have been used. To evaluate the recommendation algorithm, the precision and recall  metrics have been used in this paper. Precision is the fraction of the top-N recommended items that are relevant. Recall is the fraction of the relevant items that are recommended.

The relevant item is the item which has been rated 7 or above by the users for our case.

Table 2. Experimental Result

| Methods | Precision |
|---|---|
| Implicit rating | 60.00% |
| Simple explicit rating | 86.66% |
| Explicit rating with Method1 | 89.00% |
| Explicit rating with Method2 | 89.00% |

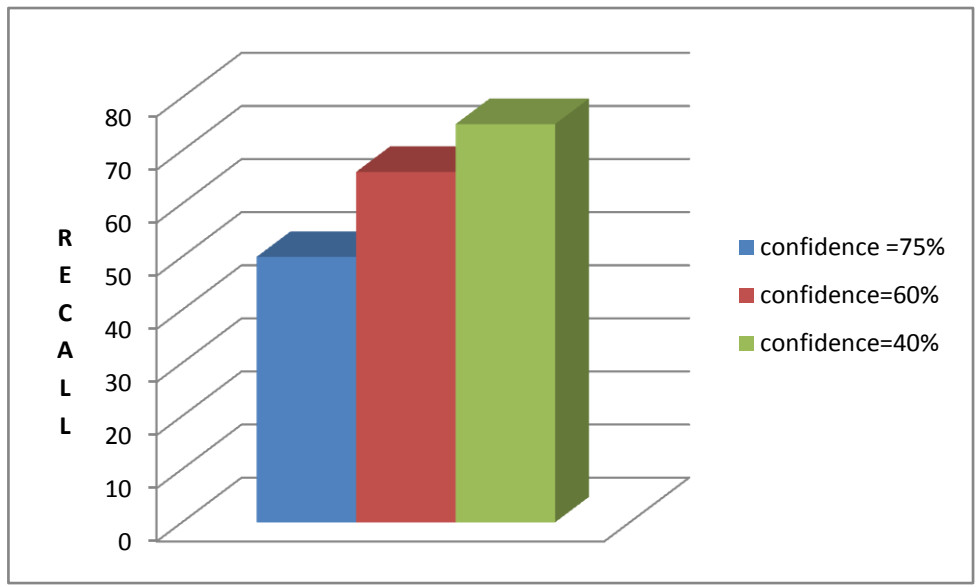

Fig. 2. Recall percentage after using Association Rule

Experimental result shown in Table 2 clearly indicates that explicit rating system is better than implicit rating system in accuracy. But for a new user, or for a user who is not interested in rating the products, implicit rating system presented in this system gives acceptable result. There is a slight improvement in explicit rating system when item frequency in transactions is considered over the case when item frequency is not considered. But in these systems though precision is good overall, recall is very low. To improve recall, Association rule mining is introduced.

## 4  Conclusions

This paper introduced weighted cosine similarity measure in e-commerce area where item frequency has been introduced. Experimental result shows that there is slight improvement in performance measure in the system presented in the paper. Recommending more than one product has been presented in this paper by using association rule which improves another performance measure 'recall' in recommending the products. This recommendation system uses transactions in making recommendation. This paper also solves the sequence recognition problem by finite state machine. It gives an acceptable result in making recommendation for a new user. In future, the performance of the same system for videos, movies can be implemented with suitable modifications.